\documentclass[twocolumn,showpacs,amssymb,prl]{revtex4}

\usepackage{graphicx}
\usepackage{dcolumn}
\usepackage{bm}

\let\vev\VEV
\def\gtap{\mathrel{ \rlap{\raise 0.511ex \hbox{$>$}}{\lower 0.511ex
   \hbox{$\sim$}}}} 
\def\ltap{\mathrel{ \rlap{\raise 0.511ex
    \hbox{$<$}}{\lower 0.511ex \hbox{$\sim$}}}} 
\newcommand{\bea}{\begin{eqnarray}} 
\newcommand{\eea}{\end{eqnarray}}
\def\beq{\begin{equation}}
\def\enq{\end{equation}}
\def\ba{\begin{eqnarray}}
\def\ea{\end{eqnarray}}
 
\def\<{<\!\!}
\def\>{\!\!>}
\def\<{\langle}
\def\>{\rangle}

\begin{document}
\input{epsf}

\title{An improved cosmological bound on the thermal axion mass}

\author{Alessandro Melchiorri$^1$, Olga Mena$^1$ and An\v{z}e Slosar$^{2,3}$}

\affiliation{$^1$ INFN Sez.\ di Roma,
Dipartimento di Fisica, Universit\`{a} di Roma``La Sapienza'', P.le
A.~Moro, 5, I-00185 Roma, Italy}
\affiliation{$^2$Astrophysics, Denys Wilkinson Building, University of Oxford, 
Keble Road, OX3RH1, Oxford, UK} 
\affiliation{$^3$Faculty of Mathematics and Physics, University of Ljubljana,
  Slovenia}

\begin{abstract}
  Relic thermal axions could play the role of an extra hot dark matter
  component in cosmological structure formation theories. By combining
  the most recent observational data we improve previous cosmological
  bounds on the axion mass $m_a$ in the so-called hadronic axion
  window.  We obtain a limit on the axion mass $m_a < 0.42$~eV at the
  $95\%$ c.l.  ($m_a < 0.72$~eV at the $99 \%$ c.l.).  A novel aspect
  of the analysis presented here is the inclusion of massive neutrinos
  and how they may affect the bound on the axion mass.  If neutrino
  masses belong to an inverted hierarchy scheme, for example, the
  above constraint is improved to $m_a < 0.38$~eV at the $95\%$ c.l.
  ($m_a < 0.67$~eV at the $99 \%$ c.l.). Future data from experiments
  as CAST will provide a direct test of the cosmological bound.
\end{abstract}

\pacs{}
\date{\today}
\maketitle

\section{Introduction}

Recent Cosmic Microwave Background and Large Scale Structure surveys
such as WMAP and SDSS have open the possibility of 
constraining fundamental physics with cosmology
(see e.g. \cite{wmap3cosm,Seljak:2006bg}).
Important upper limits on neutrino masses and 
energy densities, for example, have been obtained
which are in some cases one order of magnitude
better than the corresponding laboratory constraints
(\cite{Seljak:2006bg,fogli06,Lesgourgues:2006nd,Dodelson:2005tp}) 
or competitive with big bang nucleosynthesis
constraints (\cite{Mangano:2006ur}).

The cosmological limits are model dependent and therefore rely on
the assumption of a theoretical model of structure formation that,
even if in agreement with current data, may need further key
ingredients to explain mysteries and inconsistencies such as dark energy. 
Moreover, for some datasets, the relevance of systematics
is still matter of debate. 

However, future laboratory experiments
will certainly test the cosmological results. The overlap of 
cosmological and laboratory limits will open a new
window of investigation and may provide evidence for new physics
and/or improve our knowledge of systematics.

It is therefore timely to constrain fundamental physics with
cosmology.
In this paper we indeed move along one of those lines of
investigation, providing new bounds on the thermal 
axion mass from cosmology. There are two possible ranges of axion masses ($\sim \mu$eV and $\sim$ eV) and, in principle, both could provide either a dominant or a sub-dominant dark matter component. Here we focus on thermal axions with  masses of $\sim$~eV. For a recent revival of the cold dark matter scenario with axions of masses $\sim \mu$eV, see Ref.~\cite{murayama2}. 
New constraints on the thermal axion mass
and couplings have recently been presented by the CAST experiment, 
which searchs for axion-like particles from the Sun which couple 
to photons~\cite{cast}. While the axion mass region probed by the CAST experiment is one order
of magnitude lower than the cosmological bound presented here,
an overlap of the two results is clearly around the corner.

Let us remind the origin of the axions. Quantum Chromodynamics (QCD) respects CP symmetry,
despite the existence of a natural, four dimensional, Lorentz and
gauge invariant operator which badly violates CP. The former extra CP
violating-term gives rise to physical observables, namely, to a
non-vanishing neutron dipole moment, $d_n$. The existing tight bound
$|d_{n}| < 3 \times 10^{-26}\ e$cm~\cite{dipole} requires the CP term contribution to be very small. Why are CP violating effects so small in QCD? Why is CP not broken in QCD? This is known as the strong CP problem.  The most convincing, and elegant, solution to the strong CP problem was provided by Peccei and Quinn~\cite{PecceiQuinn}, by adding a new global $U(1)_{PQ}$ symmetry. This symmetry is spontaneously broken at a large energy scale $f_a$, generating a new spinless particle, the axion, allowing for a dynamical restoration of the CP symmetry. Axions are the pseudo Nambu-Goldstone bosons of the broken $U(1)_{PQ}$ symmetry~\cite{weinberg,wilczek} and may be copiously produced in the early universe, either thermally~\cite{turner} or non-thermally~\cite{kolb}, providing a possible (sub)dominant (hot) dark matter candidate. 
The axion mass and couplings are inversely proportional to the axion coupling constant $f_{a}$ 
\bea
 m_a = \frac{f_\pi m_\pi}{  f_a  } \frac{\sqrt{R}}{1 + R}=
0.6\ {\rm eV}\ \frac{10^7\, {\rm GeV}}{f_a}~,
\label{eq:massaxion}
\eea
where $R=0.553 \pm 0.043 $ is the up-to-down quark masses ratio~\cite{ratios} and $f_\pi = 93$ MeV is the pion decay constant.
In principle, axions can interact with photons, electrons and hadrons. If axions couple to photons and electrons, the simplest bound comes from an energy loss argument. The axions produced in a star escape carrying away energy, producing anomalous stellar observables, see Refs.~\cite{raffelt,raffelt1bis,raffelt1} for a review. However, in practice, axion interactions are model dependent. Here we
focus on \emph{hadronic axion models} such as the KSVZ
model~\cite{ksvz1,ksvz2}, in which there is no tree level interaction
between axions and leptons and the axion-photon coupling could accidentally be negligibly small. Hannestad \textit{et al}~\cite{raffelt2} have recently found an upper limit on the hadronic axion mass $m_a < 1.05$ eV ($95\%$ CL), which translates into $f_a > 5.7 \times 10^{6}$ GeV. In this \textit{letter}, we reinforce the former limit by means of an updated analysis, using a broad set of the most recent available cosmological data, and allowing for two possible hot dark matter components: neutrinos and axions.

\section{The hadronic axion model}
 Among axion couplings with hadrons, those of interest for us are the axion-nucleon couplings $\mathcal{L}_{a N}$, responsible for the processes $N + N \leftrightarrow N + N +a$ and $N+ \pi \leftrightarrow N +a $, and the axion-pion couplings $\mathcal{L}_{a \pi}$, responsible  for $a +\pi \leftrightarrow \pi + \pi$.  In practice, nucleons are so rare in the early universe respect to pions, that only the axion-pion interaction will be relevant for thermalization purposes. The lagrangian reads~\cite{chang}
\bea
{\cal L}_{a\pi} =  C_{a\pi}\frac{\partial_{\mu} a}{f_af_{\pi}}
(\pi^0\pi^+\partial_{\mu}\pi^-
+\pi^0\pi^-\partial_{\mu}\pi^+-2\pi^+\pi^-\partial_{\mu}\pi^0)~,
\label{eq:lagrangian}
\eea
where 
\bea
C_{a\pi} = \frac{1-R}{3(1+R)}
 \eea
is the axion-pion coupling constant~\cite{chang}. The most stringent limits on the axion-nucleon coupling in hadronic axion models, $g_{aN}= C_{N}m_{N}/f_a$, are those coming from SN 1987A neutrino data. If axions couple to nucleons strongly, the supernova cooling process is modified, distorting both the measured neutrino flux and the duration time of the neutrino burst emitted. The limit in the axion-nucleon coupling $g_{aN}$, assuming that the model-dependent parameter $C_N\simeq \mathcal{O}(1)$, translates into an axion decay constant $f_a \lesssim$ few $\times 10^{-6}$ GeV~\cite{pdg}. Even if axion emission does not affect the SN cooling, if $g_{aN}$ is strong enough, the axion flux may excite $^{16}O$ nuclei in water Cherenkov detectors. The absence of a large signal from radiative decays of excited $^{16}O^\star$ nuclei in the Kamiokande experiment provides a lower limit  $f_a \agt 3 \times 10^{5}$ GeV~\cite{engel}.
In summary, hadronic axions with the decay constant $f_a$ around $10^6$ GeV, i.e. $m_a\sim $ eV, can escape from all astrophysical and laboratory constraints known so far, suggesting an ideal hot dark matter candidate, within the mixed hot dark matter scenario~\cite{murayama}. 

\section{Axion decoupling}

Axions will remain in thermal equilibrium until the expansion rate of the universe, given by the Hubble parameter $H(T)$, becomes larger than their thermally averaged interaction rate. To compute the axion decoupling temperature $T_D$ we follow the usual freeze out condition
\bea
\Gamma (T_D) = H (T_D)~.
\label{eq:freezeout}
\eea 
The axion interaction rate $\Gamma$ is given by~\cite{chang}
\bea
\Gamma=n_a^{-1}\sum_{i,j} n_i n_j \vev{\sigma_{ij} v}~,
\label{eq:thermally}
\eea
where $n_a=(\zeta_{3}/\pi^2) T^3$ is the number density for axions in thermal equilibrium, and the sum extends to all production processes involving as initial states the particles $i$ and $j$, which are in equilibrium at $T_D$. 
We will assume that the axion decay constant $f_a$ is sufficiently small to ensure that axions decouple from the thermal plasma after the QCD transition epoch at $T= T_{QCD}\simeq 200$ MeV ($f_a \lesssim 4 \times 10^{7}$ GeV, i.e., $m_a \agt 0.14$ eV). Consequently, we do not have to consider axion interactions with the quarks and gluons before the QCD phase transition and the dominant processes contributing to the thermally averaged cross section in Eq.~(\ref{eq:thermally}) will be $\pi^{0} \pi^{\pm} \rightarrow a \pi^{\pm}$ and $\pi^{+} \pi^{-} \rightarrow a \pi^{0}$, see the interaction lagrangian, Eq.~(\ref{eq:lagrangian}).  We follow here the computation carried out by Chang and Choi~\cite{chang} for the average rate $\pi + \pi \rightarrow \pi +a$:
\bea
\Gamma = \frac{3}{1024\pi^5}\frac{1}{f_a^2f_{\pi}^2}C_{a\pi}^2 I~,
\eea
where 
\bea
I &=&n_a^{-1}T^8\int dx_1dx_2\frac{x_1^2x_2^2}{y_1y_2}
f(y_1)f(y_2) \nonumber \\
&\times&\int^{1}_{-1}
d\omega\frac{(s-m_{\pi}^2)^3(5s-2m_{\pi}^2)}{s^2T^4}~.
\eea
Here $f(y)=1/(e^y-1)$ denotes the pion distribution function, $x_i=|\vec{p}_i|/T$,  $y_i=E_i/T$ ($i=1,2$), $s=2(m_{\pi}^2+T^2(y_1y_2-x_1x_2\omega))$, and we assume a common mass for the charged and neutral pions, $m_\pi=138$ MeV. 

The RHS in Eq.~(\ref{eq:freezeout}) contains the Hubble expansion rate, related to the energy density of the universe via the Friedmann equation~\cite{kolb}:
\bea
H(T)=\sqrt{\frac{4 \pi^3}{45} g_\star(T)} \frac{T^2}{M_{pl}}~,
\eea
where $M_{pl}$ is the Planck mass. We have computed, for temperatures $T$ in the range $1$ MeV $< T < 200$ MeV, i.e. between BBN and the QCD phase transition eras, the number of relativistic degrees of freedom $g_\star(T)$, according to Ref.~\cite{kolb}. We neglect the axion contribution to $g_\star$ for simplicity. 
After resolving the freeze out equation Eq.~(\ref{eq:freezeout}), we obtain the axion decoupling temperature $T_D$ versus the axion mass $m_a$ (or, equivalently, versus the axion decay constant $f_a$). From the axion decoupling temperature, we can compute the current axion number density, related to the present photon density $n_\gamma=410.5 \pm 0.5$ cm$^{-3}$~\cite{pdg} via 
\bea
n_a=\frac{g_{\star S}(T_0)}{g_{\star S}(T_D)} \times \frac{n_\gamma}{2}~, 
\label{eq:numberdens}
\eea  
where $g_{\star S}$ refers to the number of \emph{entropic} degrees of freedom. 
Before electron-positron annihilation at temperatures $\sim$ eV, the number of \emph{entropic} degrees of freedom is $g_{\star S}=g_\star$,
since all relativistic particles are at the same temperature. At the current temperature, $g_{\star S}(T_0) = 3.91$~\cite{kolb}.

\section{Cosmological Constraints}

As now common practice in the literature we derive our constraints by
analyzing Monte Carlo Markov Chain of cosmological models.  For this
purpose we use a modified version of the publicly available Cosmo-MCMC
package \texttt{cosmomc} \cite{Lewis:2002ah} with a convergence
diagnostics done through the Gelman and Rubin statistic. We sample the
following eight-dimensional set of cosmological parameters, adopting
flat priors on them: the baryon and Cold Dark Matter densities,
$\omega_b=\Omega_bh^2$ and $\omega_c=\Omega_ch^2$, the ratio of the
sound horizon to the angular diameter distance at decoupling,
$\theta_s$, the scalar spectral index $n_S$, the overall normalization
of the spectrum $A$ at $k=0.05$ Mpc$^{-1}$, the optical depth to
reionization, $\tau$, the energy density in massive neutrinos

\begin{equation}
\Omega_{\nu}h^2 ={\sum m_{\nu}
\over {92.5\  \textrm{eV}}}~,
\end{equation}

\noindent  and the energy density in the 
thermal axions:

\begin{equation}
\Omega_{a}h^2 = { m_{a} n_{a}\over {1.054\cdot 10^{4}\  \textrm{eV cm}^{-3}}} = { m_{a}
\over {131\  \textrm{eV}}} \left(\frac{10}{g_{\star S}(T_D)}\right)~, 
\end{equation}
where we have used Eq.~(\ref{eq:numberdens}). For instance, for the hadronic axion upper mass bound quoted in Ref.~\cite{raffelt2}, i.e. $m_a\sim 1.05$~eV, the axion decouples at $T_{D}\sim 64$~MeV, at which $g_{\star S}(T_D)\simeq 15.24$ and $\Omega_{a}h^2 \simeq 0.0053$.

We consider a combination of cosmological data which includes the
three-year WMAP data \cite{wmap3cosm}, the small-scale CMB
measurements of CBI \cite{2004ApJ...609..498R}, VSA
\cite{2004MNRAS.353..732D}, ACBAR \cite{2002AAS...20114004K} and
BOOMERANG-2k2 \cite{2005astro.ph..7503M}. In addition to the CMB data,
we include the constraints on the real-space power spectrum of red
luminous giant (LRG) galaxies from the fourth data release of the
SLOAN galaxy redshift survey (SDSS) \cite{2004ApJ...606..702T} and 2dF
\cite{2005MNRAS.362..505C}, and the Supernovae Legacy Survey data from
\cite{2006A&A...447...31A}.  Finally we include a prior on the Hubble
parameter from the Hubble Space Telescope Key project \cite{freedman}
and the BBN prior in form of a Gaussian prior on $\Omega_b h^2$
(see e.g. \cite{Mangano:2006ur}). We refer to this dataset 
as {\it Conservative} in the rest of the paper.

In the second dataset we include constraints on the small scale linear
power spectrum coming from Lyman-$\alpha$ analysis of SDSS quasar
spectra \cite{McDonald:2004eu,McDonald:2004xn}.

\begin{table}
\begin{tabular}{ccc}
dataset/prior & $m_a<$ & $\sum m_\nu<$ \\
\hline
Conservative & 1.4 / 2.0 eV & 0.55 / 0.9 eV\\
\hline
Conservative+LYA & 0.42 / 0.72 eV& 0.20 / 0.37 eV \\
+$\Sigma m_\nu>0.05$eV & 0.41 / 0.71eV & 0.22 / 0.38eV  \\
+$\Sigma m_\nu>0.1$eV & 0.38 / 0.67eV & 0.25 / 0.44eV \\
\hline
Conservative+LYA+BAO & 0.35 / 0.64 eV& 0.18 / 0.31 eV \\

\end{tabular}
\caption{\label{tabl:1} This figure shows the 95\%/99.9\% upper
  confidence limits on the \emph{marginalised} posterior probabilities
for axion and neutrino masses. See text for discussion.} 
\end{table}

The main results of our analysis are reported in the Table
\ref{tabl:1}.  As can see, without assuming any prior on the neutrino
mass, the mass of the thermal axion is found to be $m_a <0.42$~eV and
the sum of the three active massive neutrinos $\sum m_{\nu}<0.20$~eV,
both at the $95\%$ c.l., i.e $\Omega_{a}h^2 < 0.0014$ and
$\Omega_{\nu}h^2 < 0.0018$. Therefore, the neutrino-axion (hot) dark
matter contribution represents a small fraction ($\lesssim 2.5\%$) of
the total CDM.  Excluding from the analysis the constraints from BAO
and Lyman-$\alpha$ cosmological datasets the former limits translate
into $m_a < 1.4$~eV and $\sum m_\nu < 0.55$ eV. The inclusion of the
Lyman-$\alpha$ data has an enormous impact on the analysis.  In the
same table we also consider the effect of adding Baryonic Acoustic
Oscillations (BAO) data detected in the Luminous Red Galaxies (LRG)
sample of the SDSS \cite{2005ApJ...633..560E} to the data. Strictly
speaking this is a statistically incorrect procedure as the
correlations with SDSS LRG power spectrum are not well understood, but
it gives the idea of the improvements that can be achieved by
including BAO constraints.

\begin{figure}[t]
\begin{center}
\begin{tabular}{cc}
\includegraphics[width=0.49\linewidth]{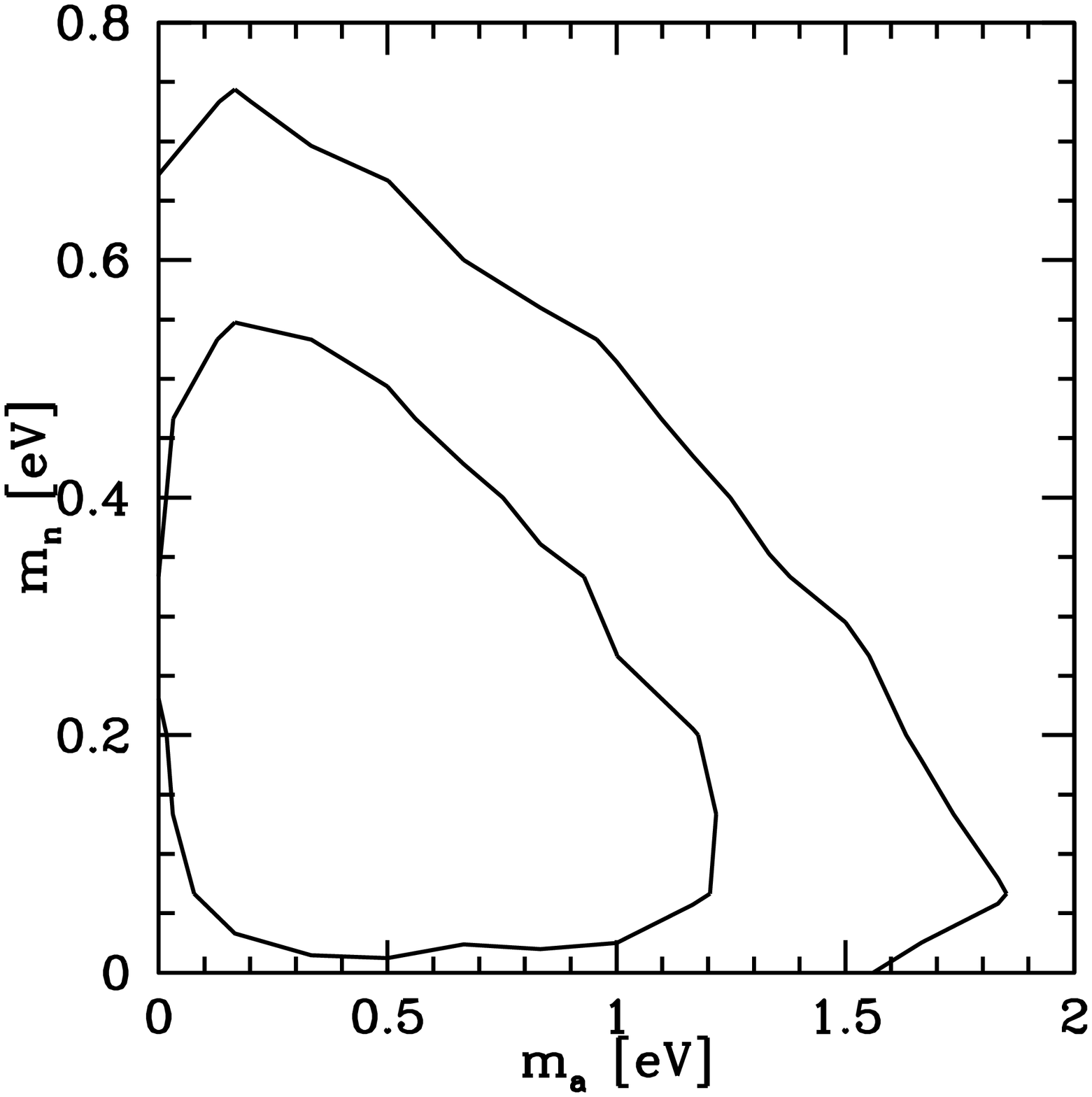} &
\includegraphics[width=0.49\linewidth]{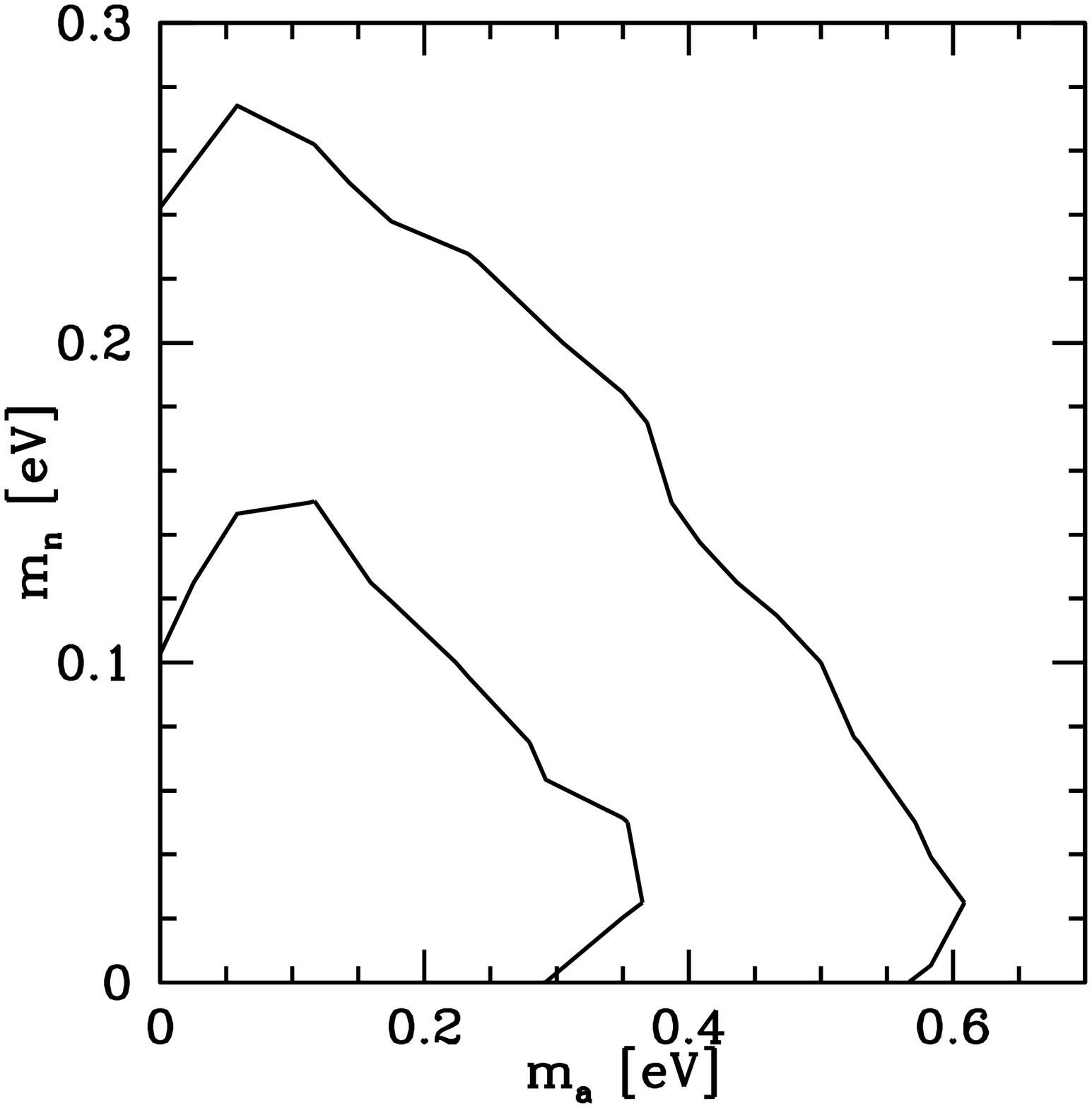} \\
\end{tabular}

\caption{Likelihood contour plot in the $\sum m_{\nu}-m_a$ 
plane showing the 68\% and 95\%  c.l.
from the {\it conservative} dataset (left panel) and
from the complete dataset (right panel). Note different axes.}
\label{fig1}
\end{center}
\end{figure}

In Fig.~\ref{fig1}, where we present marginalized constraints on the
$\sum m_{\nu} - m_a$ plane. There is a clear anti-correlation between
the constraints on the thermal axion mass and the mass of the three
active neutrinos. In other words, the cosmological data allow only for a
very specific quantity of hot dark matter: if one increases the active
neutrino mass, more hot dark matter is present in the model and the
axion mass has to be smaller in order to fit the observations.

\begin{figure}[t]
\begin{center}
\includegraphics[width=3in]{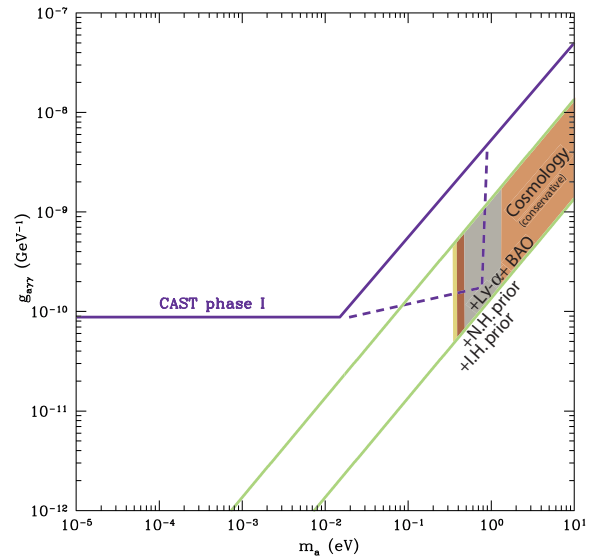}
\caption{$95\%$ CL limits on the axion mass obtained in the {\it
  conservative} and full analysis (shaded regions), assuming three possible values of the sum
  of the neutrino masses in the $m_a$-$g_{a \gamma \gamma}$ plane. 
From right to left the region represent the exclusion limits assuming
  a prior $\sum m_\nu>0$, $\sum m_\nu>0.05$ eV (N.H.) and $\sum m_\nu>0.1$ eV (I.H).
As a comparison, we show the recent results from the CAST experiment (blue contour) as well as the theoretical KSVZ parameter region (within the green lines), following Fig.~8 from Ref.~\cite{cast}, and the CAST prospects (blue dashed line)~\cite{castfuture}.}
\label{fig3}
\end{center}
\end{figure}

Figure~\ref{fig3} depicts the $95\%$ CL axion mass limits in the
$m_a$-$g_{a \gamma \gamma}$ (axion-to-photon coupling) plane. The
limits should be within the region allowed by the KSVZ model. We have
considered two possible scenarios, accordingly to neutrino oscillation
data: normal hierarchy ($\sum m_\nu \agt\sqrt{|\Delta m^2_{13}|}\agt
0.05$~eV) and inverted hierarchy ($\sum m_\nu \agt 2\cdot\sqrt{|\Delta
  m^2_{13}|}\agt 0.1$~eV), as well as the massless neutrino case. The
$95\%$ c.l. constraints that we obtain for the axion mass within the
three possible scenarios mentioned above are $m_a<0.34, 0.31$ and
$0.34$~eV, respectively, including both BAO and Lyman-$\alpha$
datasets. We found no significant difference between the normal
hierarchy and the massless neutrino scenarios. If future cosmological
data~\cite{yvonne} or direct terrestrial searches for neutrino masses,
as the ones which will be carried out by the KATRIN
experiment~\cite{KATRIN}, improve the current limits on $\sum m_\nu$,
one could obtain automatically a rather robust, independent, albeit
\emph{indirect} limit on the axion mass $m_a$.  We depict in
Fig.~\ref{fig3} the current $95\%$ c.l. CAST limit for
comparison~\cite{cast}. The CAST experiment has been upgraded and in
the near future it will explore QCD axions, that is, a range of axion
masses up to about $1$~eV. Cosmology-independent future limits on the
axion mass are therefore extremely important, since they could provide
a test of the cosmological constraint and be translated into a limit
of the universe's hot dark matter fraction in the form of massive
neutrinos.

\section{Conclusions}
We have presented an improved limit on the hadronic axion mass
by combining the most recent available cosmological data. A novel
content of this analysis is the 
addition of a hot dark matter component in the form of massive neutrinos. 
Interestingly, we have noticed an anti-correlation between the thermal
axion mass and the mass of the three active neutrinos $\sum m_\nu$. 
This anti-correlation is due to the suppression induced on the small
scale power spectra by both the relic axion and 
the massive neutrino free-streaming species. A larger (smaller) axion mass content can  be traded by a smaller (larger) massive neutrino content.
If the complete cosmological dataset is used, we find $m_a < 0.35$~eV
and  $\sum m_\nu< 0.17$~eV at the $95\%$ c.l., 
implying that the fraction of (hot) dark matter in the form of massive
thermal axions and neutrinos is only 
a few percent ($\lesssim 2.5\%$) of the total CDM content. The former
limits get modified if priors 
on the neutrino or axion masses are imposed. Future cosmological
and/or terrestrial 
searches for neutrino (axion) masses could therefore be translated 
into an improved and independent axion (neutrino) mass limit.

\section{Acknowledgments}
It is a pleasure to thank Alessandro Mirizzi for useful discussions.
The work of O.~M is supported the European Programme ``The Quest for
Unification'' contract MRTN-CT-2004-503369. Results were computed on
the UK-CCC COSMOS supercomputer.

\end{document}